\documentclass[superscriptaddress,aps,prx,reprint,twocolumn,amsmath,amssymb]{revtex4}
\usepackage{graphicx}
\usepackage{hyperref}
\usepackage{amssymb}
\usepackage{slashed}
\usepackage{dcolumn}
\usepackage{amsmath}
\usepackage{bm}
\usepackage{colordvi}
\usepackage{algorithm}
\usepackage{algpseudocode}
\usepackage{multirow}
\usepackage{titlesec}
\usepackage{newtxtext,newtxmath}
\usepackage{dsfont}
\usepackage{xcolor}
\usepackage{mathbbol}

\usepackage{mathrsfs}
\makeatletter

\newcommand{\Rmnum}[1]{\expandafter\@slowromancap\romannumeral #1@}
\makeatother

\begin{document}

\title{Terahertz-induced second-harmonic generation in quantum paraelectrics: hot-phonon effect}

\author{F. Yang}

\affiliation{Department of Materials Science and Engineering and Materials Research Institute, The Pennsylvania State University, University Park, PA 16802, USA}

\author{X. J. Li}

\affiliation{Department of Materials Science and Engineering and Materials Research Institute, The Pennsylvania State University, University Park, PA 16802, USA}

\author{D. Talbayev}
\email{dtalbayev@gmail.com}
\affiliation{Department of Physics and Engineering Physics, Tulane University, New Orleans, Louisiana 70118, USA}

\author{L. Q. Chen}
\email{lqc3@psu.edu}

\affiliation{Department of Materials Science and Engineering and Materials Research Institute, The Pennsylvania State University, University Park, PA 16802, USA}

\date{\today}

\begin{abstract} 

  Recent THz-pump second-harmonic-generation(SHG)-probe measurements of quantum paraelectrics observed a significant long-lived non-oscillatory SHG component following an ultrafast resonant excitation of the soft mode, which was interpreted as a signature of THz-induced transient ferroelectric order. {{We propose that the THz-induced modulation of the SHG signal can be attributed solely to the dynamic variation of the dielectric environment associated with the lattice background, which reflects the  coherent response of soft mode under THz pumping.}} We develop a temperature-dependent dynamic model incorporating the hot-phonon effect to simulate the soft-mode behaviors under ultrafast THz excitation.  Its application to paraelectric KTaO$_3$ produces quantitatively most of the features exhibited in our time-resolved SHG measurements and  those in existing literature, including a long-lived non-oscillatory SHG response, SHG oscillations at twice the soft-mode frequency, SHG dampings as well as temperature and field-strength  dependencies.  We conclude that the observed THz-induced non-oscillatory SHG response in quantum paraelectrics is a consequence of the  nonequilibrium hot-phonon effect, offering an alternative to its existing interpretation as a signature of transient ferroelectric order. 
  
\end{abstract}


\maketitle  

{\sl Introduction.---}The quantum criticality in condensed matter physics describes the ordering of a quantum phase that occurs near zero temperature. This phenomenon has attracted considerable attention,  due to its distinct characteristics arising from the low-lying collective excitations.  Extensive research over past few decades has suggested the presence of quantum criticality in strongly-correlated materials, which are often complex with various intertwined quantum orders. A notable exception is the displacive quantum paraelectrics~\cite{muller1979srti, rowley2014ferroelectric, cowley1996phase, fujishita2016quantum, verdi2023quantum, yang2024microscopic,singh1996stability,wu2022large}, where a strong competition between quantum fluctuations and ferroelectric ordering exists. This class of material is supposed to transition from the paraelectric to ferroelectric states at low temperature due to lattice dynamical instability, but 
the
zero-point lattice vibrations 
prohibits the long-range ferroelectric order~\cite{muller1979srti,verdi2023quantum,yang2024microscopic,wu2022large},  leading to an incipient ferroelectricity, sometimes referred to as hidden ferroelectric phase~\cite{li2019terahertz,cheng2023terahertz}. 

While doping~\cite{andrews1985x,toulouse1992precursor,aktas2014polar,rischau2017ferroelectric} or isotope substitution~\cite{takesada2006perfect,rowley2014ferroelectric} can turn quantum paraelectrics to ferroelectrics, transiently  reaching the hidden ferroelectric phase through ultrafast manipulation with intense femtosecond-pulsed laser~\cite{de2021colloquium,fechner2024quenched,basini2024terahertz,nova2019metastable} is particularly appealing~\cite{PhysRevLett.126.027602, PhysRevB.105.L100101}.  One strategy is to coherently drive the so-called soft-phonon mode---associated with lattice dynamical instability~\cite{yamada1969neutron,fleury1968electric,sirenko2000soft,shirane1967temperature}---into the nonlinear regime, aiming to create transient ferroelectric order. Several terahertz(THz)-pumped second-harmonic-generation(SHG)-probe  measurements have therefore been performed in typical displacive quantum paraelectrics, SrTiO$_3$~\cite{li2019terahertz} and KTaO$_3$ single crystals~\cite{cheng2023terahertz,li2023terahertz}. At low temperatures, a significant long-lived non-oscillatory SHG component, superimposed by a clear SHG oscillation at twice the soft-mode frequency, was observed after a THz pulse~\cite{li2019terahertz,cheng2023terahertz,li2023terahertz}. The origin and precise mechanisms of this non-oscillatory component and  soft-mode frequency doubling phenomenon~\cite{li2019terahertz,cheng2023terahertz,li2023terahertz} remain unclear~\cite{zhuang2023light,shin2022simulating}. The observed non-oscillatory SHG background is commonly interpreted as a signature of a THz-induced transient ferroelectric order,
possibly arising from a THz-driven intrinsic lattice displacement~\cite{li2019terahertz}, or from a THz-induced long-range correlation between extrinsic local polar structures by defects~\cite{cheng2023terahertz}.  However, such an interpretation seems to contradict several other findings. For example, it was  realized that the THz pulses up to 500~kV/cm were insufficient to produce a global intrinsic ferroelectricity~\cite{cheng2023terahertz}, whereas inducing a long-range correlation between extrinsic
defect dipoles via an ultrafast manipulation does not require coherently driving an intrinsic soft mode.
 More importantly, soft modes are experimentally observed to go through hardening with an increasing THz-field strength~\cite{li2019terahertz,cheng2023terahertz,li2023terahertz}, suggesting an intense THz field drives the quantum paraelectrics away from rather than towards 
ferroelectricity, since the soft-mode hardening in displacive paraelectrics is an indicator of departure from ferroelectricity.

 On the other hand, similar non-oscillatory components in pump-probe measurements have been commonly observed in other subfields, e.g., in measurements of the interband transitions in semiconductors (e.g., graphene~\cite{hafez2018extremely}) and collective excitations in superconductors~\cite{matsunaga2013higgs,matsunaga2014light,shimano2020higgs,yang2018gauge,cui2019impact}, and they were attributed to the induced nonequilibrium hot-quasiparticle effect. 
 For example, in superconductors, by using intense THz pulses, one can resonantly excite the amplitude mode of the superconducting order parameter~\cite{matsunaga2013higgs,matsunaga2014light,shimano2020higgs,yang2019gauge,yang2023optical}, namely the Higgs-mode excitation~\cite{pekker2015amplitude}. In most of the measurements, a non-oscillatory  component was observed and can persist for a long time after the THz pulse~\cite{matsunaga2013higgs,matsunaga2014light,shimano2020higgs}, as a consequence of the  nonequilibrium hot-quasiparticle effect~\cite{matsunaga2013higgs,matsunaga2014light,shimano2020higgs,yang2018gauge,cui2019impact}. {{The hot-quasiparticle effect is a fundamental aspect of nonequilibrium dynamics, applicable to both fermions (electrons) and bosons (phonons) under an external excitation due to the energy input. However, a dedicated hot-phonon effect for purely soft-phononic dynamics under ultrafast optical excitation in ferroelectrics and paraelectrics has largely been overlooked.}}

 In this work, {{we propose that a finite probe-field SHG signal
  originates from the oxygen-vacancy defects and remains nearly unchanged during the THz excitation. The observed THz-induced SHG modulation can be attributed solely to the dynamic variation of the dielectric environment associated with the lattice background, which reflects the soft mode’s coherent response under THz pumping.}} For this purpose, we develop a temperature-dependent dynamic  model to describe the soft-mode behaviors under an ultrafast resonant excitation by incorporating the hot-phonon effect, and examine the THz-induced SHG responses in quantum paraelectric KTaO$_3$ using a combination of theoretical and experimental studies.  
 
 Simulations based on our model quantitatively produce all the main features exhibited in our time-resolved SHG measurements,  including a long-lived non-oscillatory response, oscillations at twice the soft-mode frequency, SHG dampings as well as temperature and field-strength dependence. We therefore attribute the THz-induced non-oscillatory SHG component in quantum paraelectrics to the nonequilibrium hot-phonon effect. Additionally, we also explore the SHG responses of a ferroelectric KTaO$_3$ after an ultrafast THz excitation to understand the individual responses of actual ferroelectric nano-regions (e.g., extrinsic local polar structures by
defects~\cite{cheng2023terahertz}). In this case, both our theoretical simulations and experimental measurements show a THz-induced long-lived SHG oscillation at the single polar-mode frequency  without any evident signature for the non-oscillatory component, in contrast to the observed THz-induced SHG response of quantum paraelectrics, suggesting that
  the previously reported resonant SHG features in quantum paraelectrics does not come from ferroelectric nano-regions. 

\begin{figure}[htb]
  {\includegraphics[width=8.8cm]{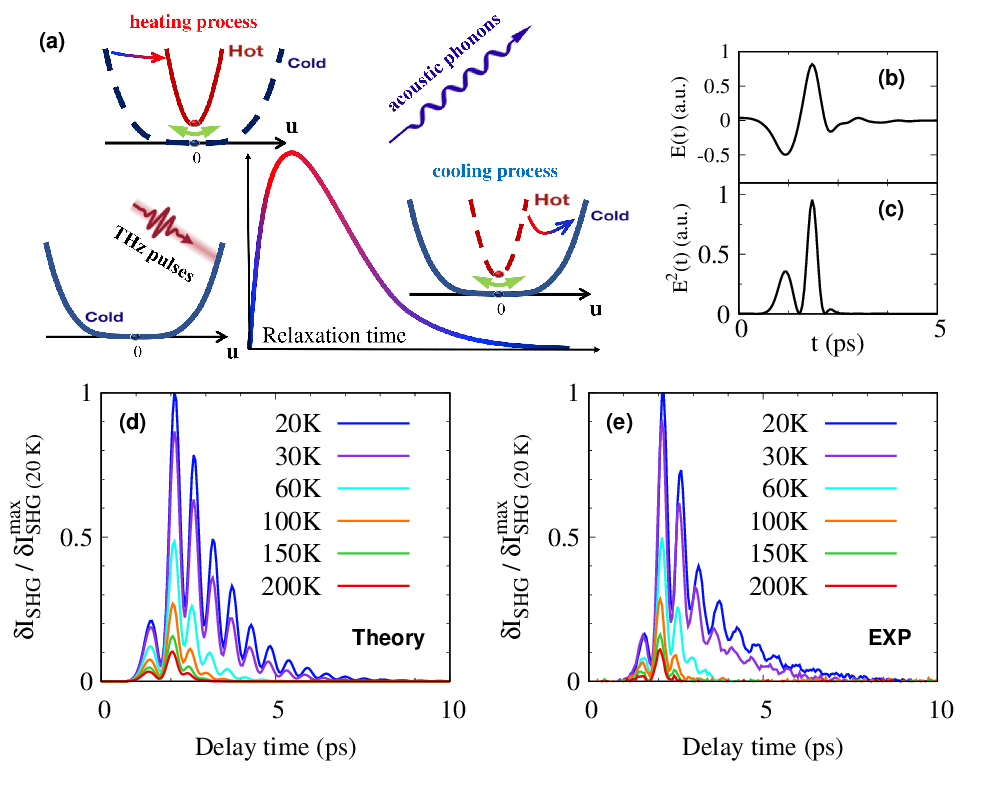}}
  \caption{{{(a) Schematic illustration of the generation (heating process) and decay (cooling process) of the hot-phonon effect during a nonlinear ultrafast resonant excitation. 
  (b) The waveform and (c) waveform squared of the single-cycle THz pump pulse employed in experiments and simulations. (d) Theoretically calculated and (e) experimentally measured time-resolved SHG in paraelectric KTaO$_3$ at different $T$. }} }    
\label{figyc1}
\end{figure}

{\sl Experimental setup.---}We use intense single-cycle THz pump pulses (up to $210~$kV/cm), generated from the optical rectification in a LiNbO$_3$ prism~\cite{li2023terahertz}, to resonantly excite the soft mode in a quantum paraelectric KTaO$_3$ single crystal and detect the time-resolved SHG signal via a femtosecond optical-frequency (800-nm) probe pulse as a function of delay time with respect to the pump pulse. The field directions of the THz pump and the optical-frequency probe pulses are both set along the [100] direction in KTaO$_3$ crystal. The time-resolved SHG signal is measured using a blue filter after the sample, which eliminates the fundamental 800-nm wavelength from the detected light. The waveform of the employed single-cycle THz pump pulses is shown in Fig.~\ref{figyc1}(b). 
 For our experimental conditions (for example, at mediate 80~K, and a laser spot size around 50~$\mu$m focused by a 200 mm lens), the SHG signal read on a lock-in amplifier detected by a GaP photodiode detector is approximately 2.9~$\mu$V with 220~$\mu$W probe intensity, confirming negligible static contribution. Under similar experimental condition but 500~$\mu$W probe intensity,  after resonant excitation by a 210~kV/cm THz field, the reading of the SHG signal reaches up to approximately 120~$\mu$V at the peak position in the positive time delay, reflecting a significant (8-times) enhancement beyond static defect-related background.

{\sl Dynamic model.---}Optical excitation of the soft mode in quantum paraelectrics is known to induce an electrical polarization ${\bf P}={u}_{\rm sp}(\sum_iQ_i{\bf e}_{i})/\Omega_{\rm cell}$~\cite{sivasubramanian2004physical,cochran1981soft,cochran1961crystal,cowley1996phase,cowley1965theory,cochran1969dynamical,cochran1960crystal,yelon1971neutron}, with ${u}_{\rm sp}$ being the soft-mode displacement; $Q_i$ and ${\bf e}_{i}$ standing for the charges and eigenvectors of the related ions (in a unit cell of volume $\Omega_{\rm cell}$) in the soft mode, respectively. As a result of lattice dynamics, its effective Lagrangian can be written as
\begin{equation}
\mathcal{L}_{\rm eff}=\frac{m_p}{2}(\partial_tP)^2-\Big[\frac{\alpha(T)}{2}P^2+\frac{b}{4}P^4-{{\bf E}(t)\cdot{\bf P}}\Big],
\end{equation}  
where $b$ is an anharmonic coefficient, related to three-phonon interactions; $m_p$ denotes the effective mass~\cite{sivasubramanian2004physical}; ${\bf E}(t)$ represents the THz field; $\alpha(T)$ is the harmonic coefficient, and using the self-consistent renormalization theory within the path-integral approach~\cite{yang2024microscopic}, the equilibrium harmonic coefficient is 
derived as $\alpha_e(T)=\alpha_e(T=0)+{b}{C}(T)$ (see Supplemental Materials~\cite{supple}). Here, $C(T)$ requires a self-consistent formulation of the bosonic thermal excitation of the soft phonons:
\begin{equation}\label{C1}
C(T)=\frac{\hbar}{m_p}\sum_{\bf q}\Big[\frac{2n^{(0)}_q+1}{\omega_q(T)}-\frac{1}{\omega_q(T=0)}\Big],
\end{equation}
where $n_q^{(0)}=1/\{\exp[\hbar\omega_q/(k_BT)]-1\}$ is the equilibrium distribution  function (Bose distribution) of the soft phonons, and the energy spectrum of the soft phonons is given by
\begin{equation}\label{omegaq}
  \omega_q(T)=\sqrt{[\alpha_e(T=0)+bC(T)]/m_p+v^2q^2},
\end{equation}
with $v$ being the mode velocity.   Mathematically,  $C(T)$ in Eq.~(\ref{C1}) increases monotonically with temperature, and as seen from Eq.~(\ref{omegaq}), this monotonic increase describes the soft-mode hardening  [i.e., the increase of soft-phonon excitation gap $\omega_{q=0}(T)$] with temperature, consistent with the known behavior of the soft modes in quantum paraelectrics~\cite{yamada1969neutron,fleury1968electric,sirenko2000soft,shirane1967temperature}.  

The ultrafast THz field ${\bf E}(t)$ can stimulate the dynamics of ${\bf P}(t)={\bf P}_0+\delta{\bf P}(t)$, and in particular, a nonequilibrium distribution $n_q=n_q^{(0)}+\delta{n_q}(t)$ of the soft phonons, thereby leading to the evolution of $\alpha(t)=\alpha_e+\delta\alpha(t)$ according to Eq.~(\ref{C1}).  Using the Euler-Lagrange equation~\cite{peskin2018introduction}, one can find the equation of motion for the polarization:
\begin{equation}\label{dtP}
m_p\partial^2_t{\bf P}+\gamma\partial_{t}{\bf P}=-\alpha(t){\bf P}-b{P^2}{\bf P}+{\bf E}(t).  
\end{equation}
Here, we have introduced a damping term, with $\gamma$ being the damping rate. This damping should be dominated by three-phonon scattering between two soft phonons and one acoustic phonon,
 leading to a $T$-dependent $\gamma(T)$. In principle, the evolution of $\alpha(t)$ should incorporate a fully microscopic bosonic Boltzmann equation of the soft phonons. However, such a microscopic treatment is complex and will not change the main results and conclusions in this work. Thus, here we employ the Allen-Cahn-like relaxation equation extensively used in the phase-field method~\cite{chen2002phase} (see  Supplemental Materials~\cite{supple}):
\begin{equation}\label{dta}
\partial_t\delta\alpha(t)=-\eta(P^2-P_0^2)/2-\delta\alpha/\tau_E,  
\end{equation}
where we take $\eta<0$ since the soft-phonon number increases after the excitation and leads to the increase of $\alpha(t)$ according to Eq.~(\ref{C1}); $\tau_E$ is the energy-relaxation time of the system. {The physical picture of hot-phonon effect is shown in Fig.~\ref{figyc1}(a). The initial THz field induces a hot-phonon effect due to the energy input, which manifests itself as the increase of the harmonic coefficient of the potential well, leading to an increased nonoscillatory SHG component in the first picosecond and hence the hardening of the soft mode oscillation (green arrow). This effect gradually decays after the THz field  due to cooling via interactions with acoustic phonons, leading to exponential decay of the non-oscillatory SHG component.}  

For the simulation in quantum paraelectrics, we first self-consistently solve the coupled Eqs.~(\ref{C1}) and~(\ref{omegaq}) using only the knowledge about the ground-state parameters $\alpha_e(T=0)$ and $b$, to obtain the equilibrium $\alpha_e(T)>0$, and hence, $P_0^2\equiv0$. We then solve the dynamic equations [Eqs.~(\ref{dtP}) and~(\ref{dta})] using experimental waveform of the single-cycle THz pump pulse [Fig.~\ref{figyc1}(b)] as the input field ${\bf E}(t)$, resulting in  the temperature-dependent soft-mode dynamics under an ultrafast excitation.

The inverse dielectric function $1/\varepsilon(T)\propto\alpha(T)$ in quantum paraelectrics~\cite{rowley2014ferroelectric,yang2024microscopic},  leading to the THz-induced SHG intensity $\delta{\rm I}_{\rm SHG}(t)\propto\delta\alpha(t)$ (See Supplemental Materials~\cite{supple}). In specific simulations, we consider the damping rate $\gamma(T)$ as the only fitting parameter for  temperature variation in the experimental measurements. Other parameters used in simulations are based on several independent experimental measurements.

{\sl Results.---}Figures~\ref{figyc1}(d) and~\ref{figyc1}(e) show the 
theoretically predicted and experimentally measured 
time-resolved SHG responses under THz excitation at different temperatures, respectively. At a low temperature of 20~K, the THz field can coherently drive the soft-mode into a strong nonlinear resonant state, showing clear oscillations on top of a non-oscillatory background after the THz stimulation ($t>2$~ps). These resonant features can persist up to 8~ps at 20~K, and gradually ebb away as temperature increases. At a high temperature of 200~K, the SHG response becomes weak and only follows the square of the THz waveform [Fig.~\ref{figyc1}(b)], because the soft mode moves out of the range of the THz spectrum due to its hardening with increase in temperature.  All of these THz-induced SHG features are consistent with the previously reported findings in quantum paraelectrics SrTiO$_3$~\cite{li2019terahertz} and KTaO$_3$~\cite{cheng2023terahertz,li2023terahertz}.

\begin{figure}[htb]
  {\includegraphics[width=8.8cm]{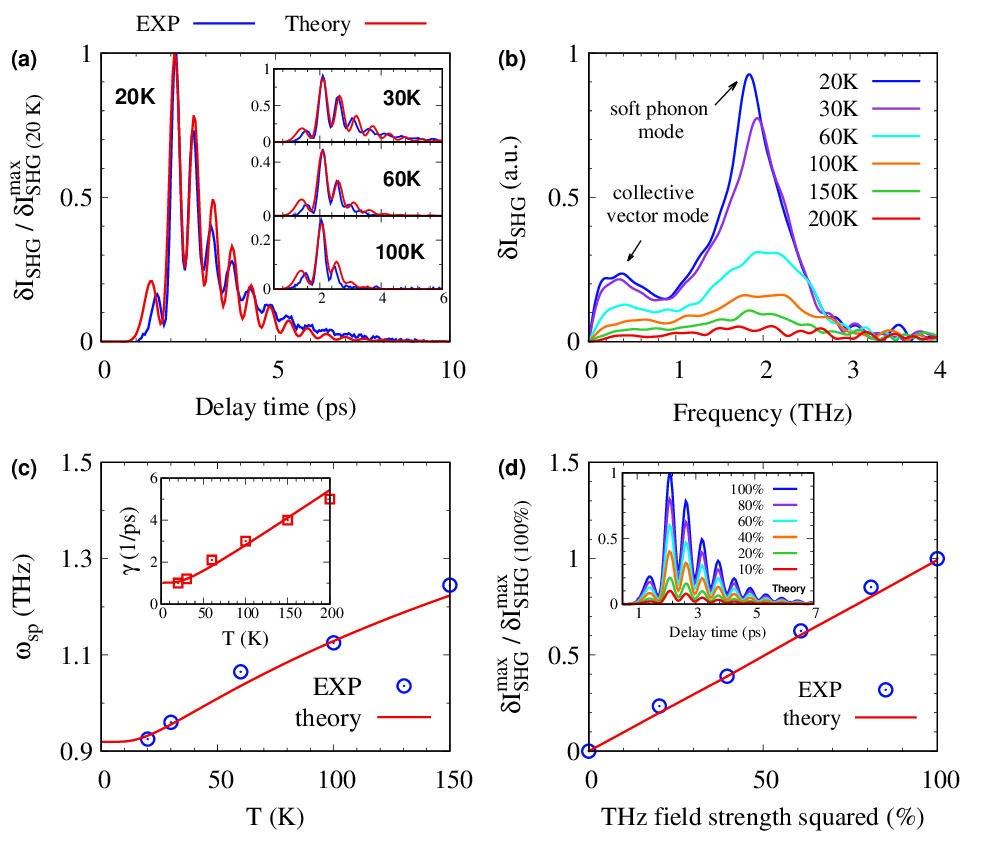}}
  \caption{{{(a) Detailed comparison between theoretical and experimental results in Fig.~\ref{figyc1}(d) and~(e). (b) Fourier transform of the measured SHG oscillations at different $T$. At temperatures below 100~K, besides the widely reported soft mode that emerges around 1.8~THz, there emerges another mode around lower frequency of 0.4~THz, which should correspond to a collective vector mode emerged in displacive ferroelectrics and quantum paraelectrics as proposed in Ref.~\cite{yang2024microscopic}. (c) Theoretical and experimental results of the soft-mode excitation gap. The inset shows the employed damping rate $\gamma(T)$ (squares) in our simulation, and a fitting curve mentioned in main text with $T_{\rm ac}=76~$K. (d) SHG peaks versus THz-field power at 20~K. The inset shows the numerical time-resolved SHG signal at different field strengths.}} }    
\label{figyc2}
\end{figure}

For a direct comparison, we plot the theoretical and experimental results together in Fig.~\ref{figyc2}(a), which demonstrates a remarkably quantitative agreement between our theoretical predictions and experimental measurements in all aspects of the THz-induced SHG responses of the quantum paraelectric.  It should be emphasized that  after the zero-temperature parameters are determined from independent measurements,  our simulation 
achieves this good agreement in both temperature variation and temporal evolution by fitting only a single parameter, the temperature-dependent damping rate $\gamma(T)$.

To gain more insight into these emerging SHG characters of quantum paraelectrics under a THz excitation, we carefully examine the temperature-dependent behaviors of the soft-mode hardening and the SHG damping. Figure~\ref{figyc2}(b) shows the FFT of the measured SHG oscillatory component, which was acquired by subtracting the non-oscillatory SHG component from the original signal. At 20~K, a single mode emerges around 1.8~THz, and it gradually hardens as temperature increases. This is consistent with the established soft-mode behavior of KTaO$_3$ reported in previous studies~\cite{cheng2023terahertz,li2023terahertz}. It should be emphasized that the soft mode in quantum paraelectric KTaO$_3$ is reported to downshift to 0.8-0.9~THz below 50~K~\cite{yamada1969neutron,fleury1968electric,sirenko2000soft,shirane1967temperature}. Thus, the THz-induced SHG oscillations in both our theory and experiment oscillate at twice the soft-mode frequency. This soft-mode frequency doubling phenomenon~\cite{li2019terahertz,cheng2023terahertz,li2023terahertz} is expected in quantum paraelectrics lying at the verge of the central-symmetric state, which we will discuss its origin later. The extracted soft-mode frequencies from experiments and the simulations [$\omega_{q=0}=\sqrt{\alpha_e(T)/m_p}$] are plotted together in Fig.~\ref{figyc2}(c) as a function of temperature and they exhibit an excellent quantitative agreement in their temperature dependence.

 For an analytical analysis of the THz excitation, we assume a pump field in the single-frequency form: ${\bf E}(t)\approx{{\bf E}_{0}}\cos{(\Omega{t})}$, with $\Omega$ being the THz-field frequency. From Eqs.~(\ref{dtP}) and~(\ref{dta}), neglecting all damping terms, one approximately has
$\partial_t\delta\alpha(t)\approx\frac{|\eta|{E}^2_{0}\cos^2(\Omega{t})}{2[m_p(\omega^2_{q=0}-\Omega^2)]^2}$  (see Supplemental Materials~\cite{supple}), leading to the THz-induced dynamic behavior of $\delta\alpha(t)$:
\begin{equation}\label{AA}
{\delta\alpha(t)}\approx\frac{|\eta|E_{0}^2\sin(2\Omega{t})/(2\Omega)}{m^2_p[(2\omega_{q=0})^2\!-\!(2\Omega)^2]^2}+\frac{|\eta|{E_{0}^2t}}{m^2_p[(2\omega_{q=0})^2\!-\!(2\Omega)^2]^2}. 
\end{equation}
The first term in the right-hand side of the above equation contributes to the SHG oscillations, induced by the second order of the THz pump field. This second-order response to a THz field leads to the aforementioned soft-mode frequency doubling phenomenon, 
since for an ultrafast ($\delta$-function) pulse, it becomes $|\eta|E_0^2\sin(2\omega_{q=0}t)/(8m_p^2\omega_{q=0}^2)$ in the response theory of a resonant excitation. The second term leads to the generation of a non-oscillatory SHG component, and it describes the nonequilibrium hot-phonon effect under the nonlinear excitation because it pushes the total $\alpha(t)$ towards a higher positive value, proportional to the power $E_0^2\sigma$ of the THz pulse, with $\sigma$ being the THz-pulse temporal width.  In addition, by Eq.~(\ref{AA}), one can also infer that a significant non-oscillatory SHG component and 
large SHG oscillations are possible only at the resonant-excitation condition of $\omega_{q=0}(T)=\Omega$, in agreement with our numerical simulations and experimental observations in the temperature-dependent SHG signal [Fig.~\ref{figyc1}(d) and~\ref{figyc1}(e)].

 Our numerical and experimental results for  the SHG signal peaks as a function of the THz-field power are plotted in Fig.~\ref{figyc2}(d). They are proportional to the THz-field power, i.e., $\delta{\rm I}_{\rm SHG}\propto{E^2_{\rm pump}E^4_{\rm probe}}$,  as also observed in  previous measurements~\cite{li2019terahertz,cheng2023terahertz,li2023terahertz} and in agreement with the analysis above. This suggests that the THz-induced SHG response in quantum paraelectrics is a second-order response to the THz field.

\begin{figure}[htb]\begin{center}
  {\includegraphics[width=7.0cm]{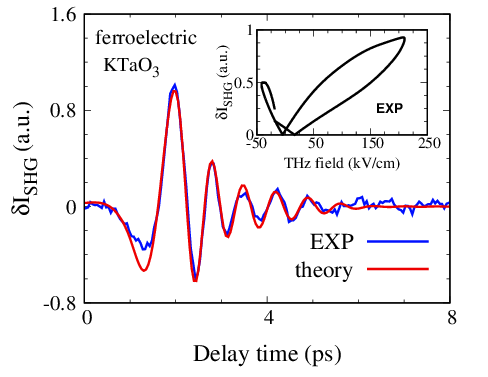}}
  \caption{{{Numerical simulation and experimental measurement of the time-resolved SHG signal in ferroelectric KTaO$_3$ at 77~K. The inset shows the measured hysteresis loop of this sample at 77~K.  }} }    
  \label{figyc3}
  \end{center}
\end{figure}

For the damping, there are two distinct relaxation processes going on after an ultrafast excitation: soft-mode damping $\gamma(T)$ and energy relaxation $1/\tau_E$. We find that the soft-mode damping dominates the damping of the THz-induced resonant SHG features as it should be, and it can be well described by the three-phonon (two soft phonons and one acoustic phonon) scattering mentioned above. Specifically, according to the microscopic scattering mechanism for calculating the scattering probabilities of the acoustic-phonon emission and absorption~\cite{yang2015hole,yang2016spin}, the damping rate of the soft mode can be approximated as $\gamma(T)\propto{2{\bar n}_{\rm ac}(T)+1}\approx{\frac{2}{e^{T_{\rm ac}/T}-1}}+1$, with ${\bar n}_{\rm ac}(T)$ being the averaged acoustic-phonon number and $T_{\rm ac}$ being a characteristic temperature. Then, the temperature dependence of $\gamma(T)$, obtained by fitting to our time-resolved experiments, can be well captured, as illustrated in the inset of Fig.~\ref{figyc2}(c).

Finally, it is noted that to explain the observed long-lived non-oscillatory SHG component, Cheng {\em et al.} has proposed a potential origin in Ref.~\cite{cheng2023terahertz}, a THz-induced correlation between the local polar structures (i.e., ferroelectric nano-regions) that arise from the extrinsic defects, leading to a global ferroelectric-like response.  To examine this possibility,  we explore the THz-induced SHG response of a ferroelectric KTaO$_3$, experimentally obtained through an annealing process~\cite{FEKTO}. We then conducted the THz-pump SHG-probe measurements on the ferroelectric KTaO$_3$ crystal, and the results are plotted in Fig.~\ref{figyc3}. As shown in Fig.~\ref{figyc3}, the ferroelectric KTaO$_3$ exhibits a totally different THz-induced SHG responses from the quantum paraelectric KTaO$_3$.  The THz-field strength dependence of the SHG response (inset of Fig.~\ref{figyc3}) shows a {{butterfly-shape hysteresis loop of the THz-induced SHG change~\cite{note2}}}, suggesting the presence of the ferroelectricity.  As for the time-resolved SHG signal (Fig.~\ref{figyc3}),  a clear oscillation develops after the THz pulse ($t>2~$ps) and persists up to 7~ps, suggesting a coherent/resonant driving of a collective excitation (i.e., a polar mode). However, no evident non-oscillatory component was observed, and no polar-mode frequency doubling phenomenon occurs in the resonant excitation as the observed SHG response during the pump pulse ($t<2$~ps) just follows the pump-pulse waveform [Fig.~\ref{figyc1}(b)]. This suggests that the THz-induced SHG signal in ferroelectric KTaO$_3$ is a linear response to the THz pump field, i.e., $\delta{\rm I}_{\rm SHG}\propto{E_{\rm pump}}{E^4_{\rm probe}}$ as a consequence of the breaking of the global lattice inversion symmetry by the existing ferroelectric order. These resonant features in the ferroelectric state are in contrast to the ones in the quantum paraelectric state, and therefore, it should suggest that the ferroelectric nano-regions are not the origin for the observed SHG resonant features in the quantum paraelectrics. 

For numerical simulations of the ferroelectric state, we set a negative value for $\alpha_e$ at $T=77~$K to fit our experimental measurement of ferroelectric KTaO$_3$, and it leads to a finite equilibrium $P_0^2=-\alpha_e/b$. Then, we perform the simulation on the basis of the dynamic model, and as shown in  Fig.~\ref{figyc3}, the produced results can well capture the experimental measurements. Following the derivation of Eq.~(\ref{AA}),  the THz-induced dynamic behavior of $\delta\alpha(t)$ in the ferroelectric KTaO$_3$ reads
\begin{equation}
  \delta\alpha\approx|\eta|\frac{({\bf P}_0\!\cdot\!{\bf E}_{0})\sin(\Omega{t})}{m_p\Omega(\Delta^2-\Omega^2)}+\frac{|\eta|E^2_{0}[\sin(2\Omega{t})/(2\Omega)+t]}{4\Omega[m_p(\Delta^2-\Omega^2)]^2},
\end{equation}
with {\small $\Delta=\sqrt{-2\alpha_e/m_p}$} being the polar-mode excitation gap.  {{The first term of this expression corresponds to the linear response of the ferroelectric order to the THz field.}}  The presence of the ferroelectric order leads to the emergence of a linear response to the THz field, and it dominates the THz-induced SHG signal at a relatively weak THz field,  in agreement with our numerical simulations and symmetry analysis as well as our observations in experimental measurements on annealed, ferroelectric KTaO$_3$ here, where the SHG intensity follows the THz pulse waveform and scales linearly with the pump field. The hot-phonon effect (non-oscillatory component) and soft-mode frequency doubling that manifest in second-order excitation (observed in quantum paraelectrics) are masked by the dominant linear response, which is why these features are not observed in our measurements on  ferroelectric KTaO$_3$, leading to the appearance of polar-mode oscillations, a linear dependence of the THz-induced SHG on pump field, and the absence of a non-oscillatory SHG component unless a significant high-order excitation by strongly intense field is present. Actually, similar phenomena have been widely observed in pump-probe measurements of other ferroelectric materials~\cite{von2018probing,grishunin2017thz,miyamoto2018ultrafast,mishina2018polarization},  including the hysteresis loop of the THz-induced SHG change~\cite{grishunin2017thz,mishina2018polarization}.

In summary, combining numerical simulations and experimental measurements on the THz-induced time-resolved SHG responses in paraelectric KTaO$_3$, we conclude that the observed long-lived non-oscillatory component in previous experiments~\cite{li2019terahertz,cheng2023terahertz} is a result of nonequilibrium hot-phonon effect. Then, the observed soft-mode hardening as THz-field strength increases~\cite{li2019terahertz,cheng2023terahertz,li2023terahertz}
can also be understood: the increase in the field strength enhances the hot-phonon effect and promotes the soft-phonon temperature during the nonequilibrium process, thereby leading to a soft-mode hardening.  

Previous experiments~\cite{li2019terahertz,cheng2023terahertz} assumes that a finite SHG
signal necessarily indicates the emergence of ferroelectric order, which breaks inversion symmetry. However, we propose and experimentally demonstrate that while a finite SHG signal indicates the
inversion symmetry breaking, it does not necessarily imply the presence of ferroelectric order.  Our measurements (refer to Supplemental Material~\cite{supple}) find that increasing oxygen vacancies in paraelectric KTaO$_3$ can significantly enhance the SHG intensity (even without the THz pump)~\cite{der1996variation}, while no signs of ferroelectric order are present in the system. {{This indicates that a finite probe-field SHG polarization can exist in quantum paraelectrics due to inevitable oxygen vacancies. It most likely arises from the two-photon inter-band transitions of electrons, mediated by the electronic defect states associated with the oxygen vacancies, which break translational symmetry and locally break inversion symmetry.}} This suggests that a finite SHG detected by the 800-nm probe field {\sl cannot} be directly used to justify the emergence of ferroelectric order. We therefore call for a careful examination of this signal.

{\sl Acknowledgments.---}F.Y. and X.J.L. did the theoretical and experimental studies, respectively, and contributed equally to this work. F.Y. and L.Q.C. acknowledge support from the US Department of Energy, Office of Science, Basic Energy Sciences, under Award Number DE-SC0020145 as part of Computational Materials Sciences Program. F.Y. and L.Q.C. also appreciate the generous support from the Donald W. Hamer Foundation through a Hamer Professorship at Penn State.
  X.J.L. and D.T. acknowledge support by the National Science Foundation under Grant No. DMR-1554866.


\end{document}


\title{Terahertz-induced second-harmonic generation in quantum paraelectrics: hot-phonon effect\\ Supplemental material}

\author{F. Yang}

\affiliation{Department of Materials Science and Engineering and Materials Research Institute, The Pennsylvania State University, University Park, PA 16802, USA}

\author{X. J. Li}

\affiliation{Department of Materials Science and Engineering and Materials Research Institute, The Pennsylvania State University, University Park, PA 16802, USA}

\author{D. Talbayev}
\email{dtalbayev@gmail.com}
\affiliation{Department of Physics and Engineering Physics, Tulane University, New Orleans, Louisiana 70118, USA}

\author{L. Q. Chen}
\email{lqc3@psu.edu}

\affiliation{Department of Materials Science and Engineering and Materials Research Institute, The Pennsylvania State University, University Park, PA 16802, USA}

\maketitle

\section{Discussion on SHG responses in quantum paraelectrics}

In this part, we discuss the SHG responses to the probe field in quantum paraelectrics and present a simple symmetry analysis of the THz-induced SHG response in both the quantum paraelectric and ferroelectric states.  In the SHG-probe measurements, the SHG intensity (as an optical-field intensity) in materials~\cite{boyd2008nonlinear,klingshirn2012semiconductor,haug2009quantum} is written as 
\begin{equation}
{\rm I}_{\rm SHG}\propto\frac{\varepsilon}{2}E^{(2\omega)}_{\rm probe}E^{(2\omega)}_{\rm probe}=\frac{1}{2\varepsilon}{p^{(2\omega)}_{\rm probe}p^{(2\omega)}_{\rm probe}},
\end{equation}
with $p^{(2\omega)}_{\rm probe}=\chi^{(2)}(\omega){E^{(\omega)}_{\rm probe}E^{(\omega)}_{\rm probe}}$ being the second-order polarization {{in the materials} generated by the 800-nm 
probe field $E^{(\omega)}_{\rm probe}$ with the high frequency of 375~THz (1.55~eV). Here, $\chi^{(2)}(\omega)$  is the second-order susceptibility.

\begin{figure}[htb]
  {\includegraphics[width=8.5cm]{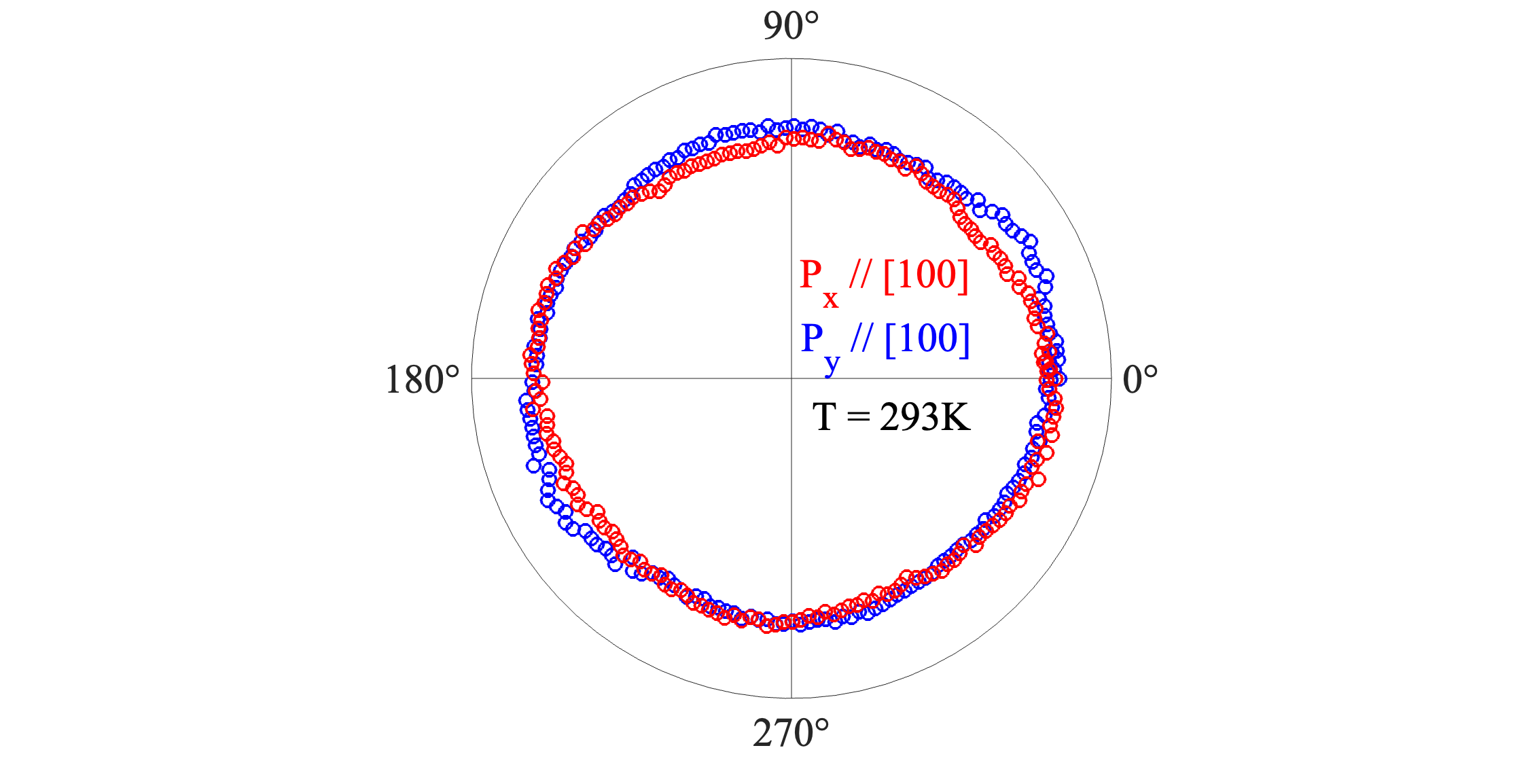}} 
  \caption{{{Room-temperature SHG polarimetry measurements of a purely oxygen-vacancy-doped KTaO$_3$ single crystal (experimentally obtained through 8-hour annealing process at 900 $^{\circ}$C under high vacuum,$10^{-5}$ Torr, without any other dopings from environment). The fundamental laser beam used here is 800-nm, which is consistent with the THz-pump optical-probe experiment.  $P_x$ and $P_y$ indicate the directions of the analyzer after the sample. The oxygen-vacancy-doped KTaO$_3$ here exhibits a strong SHG response compared to the pristine sample, at room temperature under the same experimental conditions. Despite the enhanced SHG signal, the polar plots for both $P_x$ and $P_y$ show angle-independent behavior, which indicates an isotropic crystal structure for cubic paraelectric phase rather than the anisoptropic structure for the ferroelectric phase with long-range order.  In other words,  increasing oxygen vacancies in paraelectric KTaO$_3$ can significantly enhance the SHG intensity, while no sign of ferroelectric order is present in the system. This result strongly suggest that a finite SHG signal excited by the 800-nm probe field cannot be directly used to justify the emergence of ferroelectric order. 
  }} }    \label{SHG_AN}
\end{figure}

Previous experiments~\cite{li2019terahertz,cheng2023terahertz} assumed that a finite SHG
signal necessarily indicates the presence of  the inversion-symmetry-breaking ferroelectric order. However, in fact,
many semiconducting materials exhibiting inversion symmetry breaking can generate SHG without being
ferroelectric. Here we propose that while a finite 800-nm-field SHG signal in quantum
paraelectrics indicates the inversion symmetry breaking, it does not necessarily imply the presence
of ferroelectric order, i.e., a finite $p^{(2\omega)}_{\rm probe}$ can exist in quantum paraelectrics even without the THz pump. In our experiments, we observe that increasing the oxygen vacancy concentration in paraelectric KTaO$_3$  significantly enhances the SHG intensity (see Fig.~\ref{SHG_AN}),  while no sign of ferroelectric order is present in the system, as seen from Fig.~\ref{SHG_AN}.   This supports the conclusion that the observed SHG arises from oxygen-vacancy-related defects, which locally break translational and inversion symmetries and thus satisfy the symmetry requirements for SHG. {{Similar observations have been previously reported~\cite{der1996variation}, where a reduction treatment of KTaO$_3$ (vacuum or H$_2$ atmosphere, $T=1000~^{\circ}$C) increased the optical SHG intensity (without THz pumping) whereas an oxidation treatment (O$_2$ atmosphere, $p=20$~bar, $T =750~^{\circ}$C) suppressed it.  These results were attributed to the local inversion symmetry-breaking defects associated with the oxygen vacancies rather than the global ferroelectric order.}} The  microscopic origin of this finite $p^{(2\omega)}_{\rm probe}$ in the quantum paraelectric KTaO$_3$ (with an indirect bandgap of 3.6~eV)   likely involves two-photon interband transitions mediated by electronic defect states associated with oxygen vacancies, as previously suggested~\cite{kapphan2005uv,shablaev1984band}. This further underscores that a finite SHG signal excited by the 800-nm probe field cannot be directly used to justify the emergence of ferroelectric order, in contrast to its interpretation in Refs.~\cite{li2019terahertz,cheng2023terahertz} as the emergence of a THz-induced transient ferroelectric order. Our findings therefore call for a more careful interpretation of probe-field SHG signals in such systems.

Consequently, with the presence of a finite second-order polarization $p^{(2\omega)}_{\rm probe}$ from the two-photon inter-band transitions of the electrons, we next elucidate the role of the THz pump field played in the THz-pump SHG-probe measurements of the  quantum paraelectrics, i.e., THz-induced SHG response. Specifically, the THz-induced SHG response occurs through a significant change in $1/\varepsilon$ by the coherent excitation of soft mode, which corresponds to a response from the lattice, and then, one has 
\begin{equation}
\delta{\rm I}_{\rm SHG}(t)\propto\frac{1}{2\varepsilon}{p^{(2\omega)}_{\rm probe}p^{(2\omega)}_{\rm probe}}\Big|_{\rm after ~THz~pump}-\frac{1}{2\varepsilon}{p^{(2\omega)}_{\rm probe}p^{(2\omega)}_{\rm probe}}\Big|^{\rm fundamental}_{\rm no~pump}.
\end{equation}
Thus, after the THz pump pulse, a significant nonequilibrium inversion dielectric function  $1/\varepsilon$, i.e., a significant dynamic variation of the dielectric environment associated with the lattice background,  is induced via the resonant excitation with the soft mode, thereby leading to the observation of the THz-induced SHG signal ${\rm I}_{\rm SHG}$. As it has been established that the inverse dielectric function $1/\varepsilon(T)\propto\alpha(T)$ in quantum paraelectrics~\cite{rowley2014ferroelectric,yang2024microscopic}, one has the THz-induced SHG intensity $\delta{\rm I}_{\rm SHG}(t)\propto\delta\alpha(t)$. In other words, the THz pump field resonantly excites the soft-mode dynamics, altering the dielectric environment/background  for the second-order polarization $p^{(2\omega)}_{\rm probe}$ of the probe field and hence leading to a THz-induced/enhanced SHG. {{It should be emphasized that this consideration can lead to the fact that using only ground-state equilibrium parameters, our model successfully reproduces most of the key features observed in our time-resolved SHG measurements, with remarkable quantatitive agreement between theoretical and experimental results.}}

{{The previous experiments~\cite{li2019terahertz,cheng2023terahertz} have prematurely assumed that 
 the ultrafast THz-pump field in quantum paraelectrics can alter ${\rm I}_{\rm SHG}$  through THz-induced significant changes in $\chi^{(2)}(\omega)$ and hence $p^{(2\omega)}_{\rm probe}$ (e.g., from zero to a large value). This is indeed a possible mechanism for THz-induced SHG.    
  However, the frequency of the 375~THz (1.55~eV) probe field is significantly high. To trigger the possible mechanism here, coherently driving the THz-frequency soft mode with a 0.9-THz pump field would need to transiently create a high-energy ($\sim3.1~$eV) collective excitation or quasiparticle in quantum paraelectrics that can respond at the second harmonic of the probe field (750~THz, 3.1~eV), or generate a transient channel for mediating the significantly high-energy  transitions.  Given the large energy mismatch, this scenario is unlikely. The hyper-Raman effect could be a possible mechanism. However, this effect typically requires very high power, while the probe field we used is very weak, making contribution from this effect nearly undetectable. Additionally, as we have mentioned above, the frequency of the 375~THz (1.55~eV) probe field is significantly higher than the soft-phonon frequency here ($\omega_{\rm op}=0.9~$THz at low temperatures). Given this large frequency mismatch, we do not expect a clear or visible hyper-Raman effect arising from the interaction of the probe field at $(2\times375)\pm0.9~$THz.  While we cannot exclude all other possible scenarios here, they must also theoretically address and justify which microscopic excitations  (electrons or phonons or other quasiparticles) can respond to the high-frequency (750~THz, 3.1~eV) field after the THz pump and how such a response to high-frequency field at the second harmonic can occur.}} \\

Last but not least, the THz-induced SHG intensity from our experimental measurements and numerical simulations in the main text as well as previous experimental measurements~\cite{li2019terahertz,cheng2023terahertz,li2023terahertz} is 
\begin{equation}
{\delta}{\rm I}_{\rm SHG}\propto{E_{\rm pump}^2E_{\rm probe}^4},  
\end{equation}
which is invariant under the inversion operation on the pump field. Nevertheless, in the ferroelectric KTaO$_3$,  the THz-induced SHG intensity from our experimental measurements and numerical simulations in the main text is
\begin{equation}\label{FEEE}
{\delta}{\rm I}_{\rm SHG}\propto{E_{\rm pump}E_{\rm probe}^4},  
\end{equation}
which is not inversion-symmetric due to the linear response to the pump field. Therefore, a finite SHG response in Eq.~(\ref{FEEE}) requires the existence of the breaking of the global {\it lattice} inversion symmetry, which can be provided by the presence of the ferroelectric order as in the first term of Eq.~(7) in the main text.

\section{Derivation of the effective equilibrium Lagrangian}

In this section, within the fundamental path-integral approach, we derive the effective equilibrium Lagrangian of the soft-mode-related polarization, in order to formulate the soft-mode hardening with temperature.

The excitation of the soft-phonon displacement in quantum paraelectrics can lead to an electrical polarization~\cite{sivasubramanian2004physical,cochran1981soft,cochran1961crystal,cowley1996phase,cowley1965theory,cochran1969dynamical,cochran1960crystal,yelon1971neutron}. Considering a polarization field ${\hat {\bf P}}$, the Lagrangian is written as~\cite{rowley2014ferroelectric,conduit2010theory,roussev2003theory,palova2009quantum,matsushita2007model,lines2001principles}
\begin{equation}\label{BL}
  \mathcal{L}_0=\frac{m_p}{2}(\partial_t{\hat P})^2-\Big[\frac{g}{2}({\nabla {\hat P}})^2+\frac{a}{2}{\hat P}^2+\frac{b}{4}{\hat P}^4\Big],
\end{equation}
as a result of lattice/phonon dynamics. Here, $a$ and $b$ are bare/ground-state model (harmonic and anharmonic) parameters; $g=v^2m_p$ is a parameter related to the velocity $v$ of the soft-phonon mode. In the renormalization theory~\cite{yang2024microscopic}, the polarization field ${\hat {\bf P}}={\bf P}+\delta{\bf P}$, consisting of the homogeneous polarization ${\bf P}$ (associated with the soft mode at $q=0$) and inhomogeneous fluctuations $\delta{\bf P}$ (associated with the soft phonons at $q\ne0$). Then,  the action in the Matsubara representation reads
\begin{equation}\label{action}
  S=\sum_{\bf q}\int_0^{\hbar\beta}d{\tau}\Big[\frac{m_p}{2}(\partial_{\tau}\delta{P})^2\!+\frac{g}{2}q^2\delta{P}^2\!+\!\frac{a}{2}{\delta}P^2\!+\!\frac{b}{4}{\delta}P^4\!+\!\frac{b}{2}{P}^2\delta{P}^2\Big]+\frac{a}{2}P^2+\frac{b}{4}P^4,~~~~
\end{equation}
where $\beta=1/(k_BT)$ and we have taken the vanishing correlations $\langle({\bf P}\cdot{\delta{\bf P}})^{2}\rangle=0$ by considering a zero soft-mode-related polarization ${\bf P}=0$ at the realistic case in quantum paraelectrics. To obtain the effective action of the homogeneous polarization, one can perform the standard integration over the bosonic field of the polarization fluctuation within the path-integral formalism. Here we present two self-consistent methods. 

{\sl Generating functional methods.---}Under the mean-field approximation, one can find an effective equilibrium Lagrangian of the soft mode from the action in Eq.~(\ref{action}): $\mathcal{L}_{\rm eff}={\alpha_e(T)}P^2/2+{b}P^4/4$ with $\alpha_e(T)=a+b\langle{\delta{P}^2}\rangle$. To calculate $\langle{\delta{P}^2}\rangle$, within the path-integral formalism, using the action in Eq.~(\ref{action}), the thermally averaged polarization fluctuation can be written as~\cite{peskin2018introduction,abrikosov2012methods}
\begin{eqnarray}
  \langle{\delta{P}^2}\rangle&=&\!\sum_{\bf q}\Big\langle\Big|\delta{P}(\tau,{\bf q})\delta{P}(\tau,-{\bf q})e^{-\frac{S}{\hbar}}\Big|\Big\rangle=\!\sum_{\bf q}\int\!\!\frac{D\delta{P}}{\mathcal{Z}_0}~\delta_{J_{\bf q}}\delta_{J_{-{\bf q}}}\big\{e^{-\frac{S}{\hbar}-\int_0^{\hbar\beta}d\tau{d}{\bf q'}[J_{\bf q'}\delta{P}(\tau,{\bf q'})]}\big\}\big|_{J=0}\nonumber\\
  &=&\!\sum_{\bf q}\int\!\!\frac{D\delta{P}}{\mathcal{Z}_0}e^{-aP^2/(2\hbar)-{b}P^4/(4\hbar)}\delta^2_{J_{\bf q}}\Big\{e^{-\int_0^{\hbar\beta}d\tau{d}{\bf q'}\big\{\frac{1}{2\hbar}\delta{P}(\tau,{\bf q'})[m_p\omega_{q'}^2(a,\langle{\delta{P}^2}\rangle)-m_p\partial_{\tau}^2]\delta{P}(\tau,{\bf q'})+J_{\bf q'}\delta{P}(\tau,{\bf q'})\big\}}\Big\}\Big|_{J=0}\nonumber\\
  &=&\!\sum_{\bf q}\delta_{J_{\bf q}}^2\exp\Big[\int_0^{\hbar\beta}d\tau{d}{\bf q'}\Big(\frac{1}{2}J_{\bf q'}\frac{\hbar/m_p}{\omega_{q'}^2(a,\langle{\delta{P}^2}\rangle)-\partial_{\tau}^2}J_{\bf q'}\Big)\Big]\Big|_{J=0}=-\!\sum_{\bf q}\frac{1}{\hbar\beta}\sum_{n}\frac{\hbar/m_p}{(i\omega_n)^2\!-\!\omega_{q}^2(a,\langle{\delta{P}^2}\rangle)}\nonumber\\
  &=&\sum_{\bf q}\frac{\hbar}{m_p}\frac{\coth\big[\hbar\beta\omega_{q}(a,\langle{\delta{P}^2}\rangle)/2\big]}{2\omega_{q}(a,\langle{\delta{P}^2}\rangle)},\label{pppp}~~~~~~
\end{eqnarray}
with
\begin{equation}
\omega_{q}^2(a,\langle{\delta{P}^2}\rangle)=(a+b\langle\delta{P}^2\rangle+b\langle{P^2}\rangle)/m_p+v^2q^2=(a+b\langle\delta{P}^2\rangle)/m_p+v^2q^2.  
\end{equation}
Here, $\omega_n=2n\pi{k_BT}/\hbar$ represents the bosonic Matsubara frequencies; $J_{\bf q}$ denotes the generating functional and $\delta{J_{\bf q}}$ stands for the functional derivative~\cite{peskin2018introduction,yang2020theory,yang2024microscopic}; $\mathcal{Z}_0=\langle|e^{-S}|\rangle$ is the normalization factor. It is noted that we utilized the mean-field approximation to obtain the soft-phonon energy spectrum, thereby leading to a self-consistent formulation of $\langle\delta{P^2}\rangle$;

{\sl Self-consistent Green function methods.---}We re-write the action in Eq.~(\ref{action}) as
\begin{equation}
  S\!=\!\sum_{\bf q}\int_0^{\hbar\beta}\!\!d{\tau}\frac{1}{2}\delta{P}(\tau,{\bf q})D^{-1}\big(\partial_{\tau},q\big)\delta{P}(\tau,{\bf q})+\frac{a}{2}P^2+\frac{b}{4}P^4,
\end{equation}
where the inverse Green function is defined as~\cite{peskin2018introduction,abrikosov2012methods}
\begin{equation}
  D^{-1}\big(\partial_{\tau},q\big)=gq^2\!+\!a\!+\!{b}{\delta}P^2/2\!+\!b{P}^2-m_p\partial_{\tau}^2.
\end{equation}
Performing the integration over the bosonic field of the fluctuation within path-integral formalism, one finds the effective action:
\begin{equation}
  S_{\rm eff}=\frac{1}{2}\hbar{\rm \bar{T}r}\ln\big[{D^{-1}(\partial_{\tau},q\big)}\big]+\frac{a}{2}P^2+\frac{b}{4}P^4.  
\end{equation}
Through the variation with respect to $P$, the equation to determine the soft-mode-related polarization is given by 
\begin{equation}
 \frac{1}{2}\hbar{\rm \bar{T}r}\Big\{\partial_{P_r}\big[{D^{-1}\big(\partial_{\tau},q\big)}\big]D(\partial_{\tau},q)\Big\}+{\partial_{P}}\Big(\frac{a}{2}P^2+\frac{b}{4}P^4\Big)=0.  
\end{equation}
Imposing the mean-field approximation, the above equation becomes
\begin{eqnarray}
  && \left\langle\hbar{\rm \bar{T}r}\Big\{\frac{b{P}}{gq^2\!+\!a\!+\!{b}{\delta}P^2/2\!+\!b{P}^2-m_p\partial_{\tau}^2}\Big\}\right\rangle\!+\!aP\!+\!bP^3\!=\!0\Rightarrow\!{\rm \bar{T}r}\Big\{\frac{\hbar\langle{bP}\delta{P}^2\rangle}{\langle(gq^2\!+\!a\!+\!{b{P}^2})\delta{P}^2+\!{b}{\delta}P^4/2\!-\!m_p\delta{P}\partial_{\tau}^2\delta{P}\rangle}\Big\}\!+\!aP\!+\!bP^3\!=\!0\nonumber\\
  &&\Rightarrow\!\!\sum_{\bf q}\frac{1}{\beta}\sum_n\frac{bP\langle\delta{P}^2\rangle}{\langle[gq^2\!+\!a\!+\!b{P}^2\!-\!m_p(i\omega_n)^2]\delta{P}^2+\!{b}{\delta}P^4/2\rangle}\!+\!aP\!+\!bP^3\!=\!0\nonumber\Rightarrow\!\!\sum_{\bf q}\frac{1}{\beta}\sum_n\frac{bP/m_p}{\omega_{q}^2(a,\langle{\delta{P}^2}\rangle)-(i\omega_n)^2}\!+\!aP\!+\!bP^3\!=\!0\nonumber\\ 
  &&\Rightarrow\!\!\big(a\!+\!b\langle\delta{P}^2\rangle\!\big){P}\!+\!b{P}^3\!=\!0,
\end{eqnarray}
with
\begin{equation}
  \langle\delta{P}^2\rangle=\!\!\sum_{\bf q}\frac{1}{\beta\hbar}\sum_{n}\frac{\hbar/m_p}{\omega_{q}^2(a,\langle{\delta{P}^2}\rangle)-(i\omega_n)^2}=\sum_{\bf q}\frac{\hbar}{m_p}\frac{\coth\big[\hbar\beta\omega_{q}(a,\langle{\delta{P}^2}\rangle)/2\big]}{2\omega_{q}(a,\langle{\delta{P}^2}\rangle)}. 
\end{equation}
Then, one finds an effective equilibrium Lagrangian $\mathcal{L}_{\rm eff}={\alpha_e(T)}P^2/2+{b}P^4/4$ with $\alpha_e(T)=a+b\langle{\delta{P}^2}\rangle$ again. 

{\sl Thermal fluctuations.---}Due to the bosonic excitation of the soft phonons, $\langle{\delta{P^2}}\rangle$ consists of the thermal fluctuations $\langle\delta{P}^2_{\rm th}(T)\rangle$ and zero-point fluctuations $\langle\delta{P}^2_{\rm zo}\rangle$. In the renormalization theory~\cite{yang2024microscopic},  the zero-point fluctuations should be integrated to the ground state, as in the quantum-field description of the vacuum. Consequently, the bare/ground-state parameter $a$ is experimentally unobservable, and only through the renormalization by zero-point fluctuations, $a$ becomes the zero-temperature parameters $\alpha_e(T=0)$ that can be experimentally measured. Further applying the renormalization by thermal fluctuations leads to the finite-temperature $\alpha(T)$. Following this description, the renormalization processes by the zero-point fluctuations read
\begin{eqnarray}\label{a0}
\alpha_e(T=0)&=&a+b\langle\delta{P}^2_{\rm zo}\rangle, 
\end{eqnarray}  
where the zero-point fluctuations:
\begin{equation}\label{zo}
 \langle{\delta{P}^2_{\rm zo}}\rangle=\int\frac{\hbar}{2m_p}\frac{1}{\omega_{q}(a,\langle{\delta{P}^2_{\rm zo}}\rangle)}\frac{d{\bf q}}{(2\pi)^3}.  
\end{equation} 
Then, the self-consistent renormalization processes by the thermal fluctuations at finite temperatures are given by
\begin{eqnarray}
  \alpha_e(T)&=&\alpha_e(T=0)+b\langle\delta{P}^2_{\rm th}(T)\rangle,\label{aaaa}  
\end{eqnarray}
where the thermal fluctuations are determined by subtracting the zero-temperature part and are written as 
\begin{equation}
\langle{\delta{P}^2_{\rm th}}\rangle=\int\frac{\hbar}{2m_p}\Bigg\{\frac{\coth\big[\hbar\beta\omega_{q}\big(\alpha_e(T=0),\langle{\delta{P}^2_{\rm th}}\rangle\big)/2\big]}{\omega_{q}\big(\alpha_e(T=0),\langle{\delta{P}^2_{\rm th}}\rangle\big)}-\frac{1}{\omega_{q}\big(\alpha_e(T=0),0\big)}\Bigg\}\frac{d{\bf q}}{(2\pi)^3}.   
\end{equation}
For accuracy, one can directly start with the experimentally measured zero-temperature parameters and perform the renormalization by thermal fluctuations to obtain the finite-temperature properties.

\section{Derivation of dynamic equations}

In this section, we present the derivation of the dynamic equations under an ultrafast excitation. Specifically, as mentioned in the main text, the external ultrafast pump field ${\bf E}(t)$ can induce the dynamic of ${\bf P}(t)$ and in particular, the dynamic of $\alpha(t)$ because of the field-induced nonequilibrium distribution $n_{\bf q}(t)$ of soft phonons in paraelectric phase or collective excitations of the polar mode in ferroelectric phase. For a general case, we assume ${\bf P}(t)={\bf P}_0+\delta{\bf P}(t)$ and $\alpha(t)=\alpha_e+\delta\alpha(t)$, with equilibrium $P_0^2=0$ for $\alpha_e>0$ representing the paraelectric phase and $P_0^2=-\alpha_e/b$ for $\alpha_e<0$ representing the ferroelectric phase. Based on the effective Lagrangian $\mathcal{L}_{\rm eff}$ of the homogeneous polarization [Eq.~(1) in the main text], using the Euler-Lagrange equation of motion with respect to ${\bf P}(t)$, i.e., $\partial_{\mu}\big[\frac{\partial\mathcal{L}_{\rm eff}}{\partial(\partial_{\mu}{\bf P})}]=\frac{\partial\mathcal{L}_{\rm eff}}{\partial{\bf P}}$, one has
\begin{equation}\label{SP}
m_p\partial^2_t{\bf P}+\gamma\partial_{t}{\bf P}=-\alpha(t){\bf P}-b{P^2}{\bf P}+{\bf E}(t),    
\end{equation}
where we have introduced a damping term $\gamma\partial_{t}\delta{\bf P}$ with $\gamma$ being the damping rate. Microscopically, this damping should arise from the three-phonon scattering between two soft phonons and one acoustic phonon as shown in Fig.~\ref{Sfigyc}.

\begin{figure}[htb]\begin{center}
  {\includegraphics[width=8.0cm]{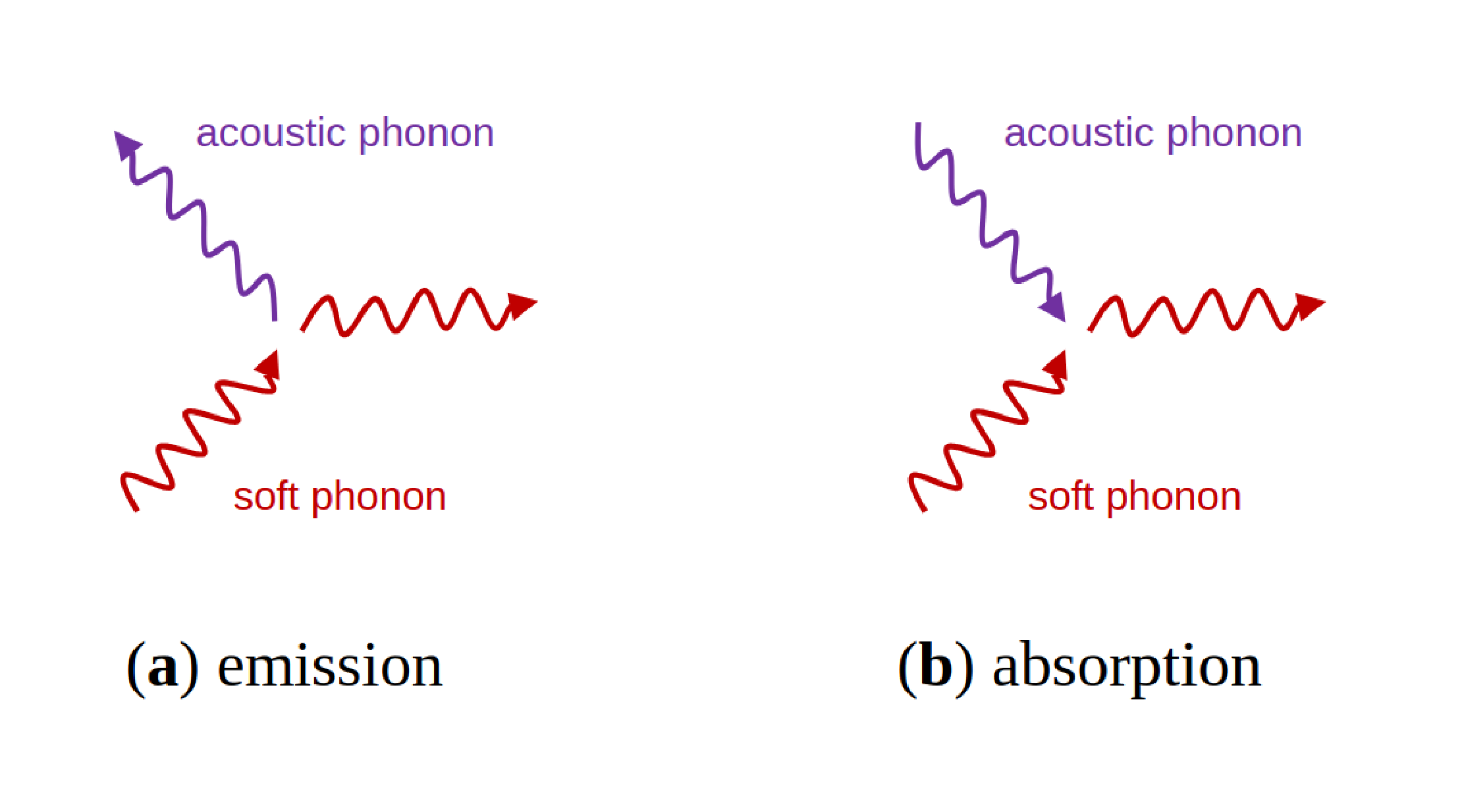}}
  \caption{Scattering processes of soft phonons, caused by three-phonon interaction between two soft phonons and one acoustic phonon.}    
  \label{Sfigyc}
  \end{center}
\end{figure}

The dynamics of $\delta\alpha(t)$ should incorporate a fully microscopic bosonic Boltzmann equation of the soft phonons, from which one in principle can find a hot-phonon distribution with hot temperature and finite chemical potential during the nonequilibrium process after the ultrafast excitation  because of the microscopic phonon-phonon scatterings/interactions.  Including the microscopic optical excitation of soft phonons in such a complex microscopic treatment  remains an open issue in the literature. Here since the field-induced nonequilibrium distribution $n_{\bf q}(t)$ of soft phonons is directly related to $\alpha(t)$ according to Eq.~(2) in the main text, we employ the Allen-Cahn-like macroscopic dynamic equation extensively used in the phase-field method~as an approximation\cite{chen2002phase}:
\begin{equation}\label{SA}
\partial_{t}\alpha(t)=-\eta\partial_{\alpha}\Big[F(t)-F(t=-\infty)\Big]-\frac{\alpha(t)-\alpha_e}{\tau_E}=-\frac{\eta}{2}(P^2-P_0^2)-\frac{\alpha(t)-\alpha_e}{\tau_E},  
\end{equation}
where $F(t)$ is the nonequilibrium free energy of the homogeneous polarization and we have introduced a relaxation (second) term by considering the relaxation-time approximation.   

{{{\sl Note added.---}In our model, we in fact assume that the relationship $1/\varepsilon\propto{\alpha}$ at equilibrium~\cite{rowley2014ferroelectric} can extend to the non-equilibrium ultrafast regime. We base this assumption on the fact that, the ultrafast dynamics on the timescales of several picoseconds in the present study should be well-approximated as a Markovian process, rather than the non-Markovian behavior that typically occurs on timescales of several femtoseconds. This assumption allows us to extend the equilibrium relationship $1/\varepsilon\propto{\alpha}$ into the nonequilibrium regime.  While this remains an approximation, it is justified by the timescale of the dynamics we are investigating and provides a reasonable foundation for our model. }}

{{Another issue concerns the dynamics of $\alpha(t)$ and hot-phonon  temperature.  As pointed out in several previous theoretical works~\cite{rowley2014ferroelectric,palova2009quantum,rechester1971contribution}, the temperature dependence of the equilibrium harmonic coefficient $\alpha_e$ can be described by considering the thermal excitation of the soft phonons. These works provide a solid foundation for understanding the temperature dependence of $\alpha$ in equilibrium. Since we assert that the ultrafast processes in the present study should be approximated as a Markovian process, rapid phonon-phonon scattering plays a crucial role in reshaping the phonon distribution that closely resembles a quasi-hot Bose-Einstein distribution. As a consequence, the concept of a time-dependent hot-phonon temperature becomes valid in this context and the dynamics of $\alpha(t)$ can be effectively captured by a time-dependent temperature. }}
    
{{It is important to clarify that while the system is coherently driven by the THz pulse, which excites the soft mode at $q\sim0$ (long wavelength), this resonant excitation at the single $q\sim0$ point does not imply that the entire phonon distribution remains coherent.  We propose that due to the rapid scattering, the phonon distribution for all $q\ne0$, i.e., the thermal-like behavior for most modes,  will evolve into a quasi-hot Bose-Einstein distribution, and hence, the system's behavior rapidly transitions into a quasi-thermalized regime, even though the system is initially driven in a coherent state by the THz pulse.
 This also justifies the use of a time-dependent temperature to describe the dynamics of $\alpha(t)$. }}

 {{If more ultrafast manipulation (e.g., using attosecond laser beams) can be experimentally performed, a more detailed analysis may reveal deviations from this thermal approximation for the first few femtoseconds. In such a scenario, the dynamics of $\alpha(t)$ might no longer be accurately described by a time-dependent temperature, and the system would need to be modeled with a more detailed treatment of the non-Markovian behavior. 
  }} 

{{Theoretically, a full microscopic treatment of the nonequilibrium soft-phonon distribution $n(t)$ would involve a detailed Boltzmann equation for the soft phonons, which would require addressing complete and full phonon-phonon scatterings and interactions for all phonon branches at all wavevectors $q$.  Moreover, regarding the coherent excitation of the soft mode at $q\sim0$, the source term for this excitation in  Boltzmann equation is very challenging to describe and address in a fully microscopic way, as it would require detailed modeling of the interaction between the optical pump and the phonon modes in Boltzmann equation. All of these then become a complex task that goes beyond  existing theoretical approaches. In this situation, we use a semi-microscopic/semi-classical approach to describe the nonequilibrium dynamics.  Specifically, we employ the microscopic Bose-Einstein distribution to capture the equilibrium harmonic coefficient, and then, employ a Allen-Cahn-like classical equation to capture the nonequilibrium soft-phonon dynamics at different temperatures. This simplification avoids the need for a fully microscopic nonequilibrium solution (e.g., solving the bosonic Boltzmann equation for all $q$). This semi-microscopic approach so far is sufficient to describe the main dynamics and provide a reasonable interpretation of  experimental results,  particularly considering the fact that our numerical calculations show that our model is more than adequate to quantitatively predict the THz-induced SHG. A full microscopic treatment, if it can be done, will not alter the main results and conclusions of the paper.}}

\begin{center}
\begin{table}[htb]
  \caption{Specific parameters used in our simulation. For the quantum paraelectric KTaO$_3$,  the model parameters at zero temperature: the velocity $v$ of the soft-phonon mode was determined in Ref.~\cite{rowley2014ferroelectric} by comparing the data from inelastic neutron~\cite{shirane1967temperature} and Raman scattering~\cite{fleury1968electric} experiments at 4~K; $\alpha_e(T=0)$ and $b$ were determined in Ref.~\cite{yang2024microscopic} by comparing the data from experimental measurement of inverse dielectric function at low temperatures~\cite{aktas2014polar}; the integral cutoff $q_c$ is taken as the one in Ref.~\cite{rowley2014ferroelectric}; $m_p$ is determined by $\Delta^2_{\rm op}(T=0)=\alpha_e(T=0)/m_p$ according to Eq.~(3) in the main text, where $\Delta_{\rm op}(T=0)$ denotes the soft-mode excitation gap at low-temperature limit; $\Delta_{\rm op}(T=0)$ and $\tau_E$ are approximately taken as the experimental values at 20~K from our time-resolved SHG measurement. As for the ferroelectric KTaO$_3$ at 77~K, only $\alpha_e$ and $\gamma$ are changed and are determined by fitting to our experimental data of the time-resolved SHG signal. We find the choice of parameter $\eta$ in simulation does not affect the normalized SHG signal presented in the figures of the main text.}  
\renewcommand\arraystretch{1.5}  \label{KTO}
\begin{tabular}{l l l l l l l}
    \hline
    \hline  
   paraelectric KTaO$_3$&\quad$\alpha_e(T=0)~$(meV$\cdot${\AA}/e$^2$)&\quad$b~$(meV$\cdot${\AA}$^5$/e$^4$)&\quad$\Delta_{\rm op}(T=0)~$(THz)&\quad$v~$({\AA}/ps)&\quad$\tau_E^{-1}$~(1/ps)&\quad$q_{c}$~({\AA}$^{-1}$) \\
    &\quad\quad$20.1\times10^{-5}/\varepsilon_0$&\quad$0.16\times(2\pi)^3/\varepsilon_0$&\quad\quad\quad0.92&\quad\quad57&\quad\quad8&\quad\quad0.134\\
    \hline
    ferroelectric KTaO$_3$&\quad$\alpha_e(T=77~K)~$(meV$\cdot${\AA}/e$^2$)&\quad$\gamma~$(1/ps)\\
    &\quad\quad$-12.94\times10^{-5}/\varepsilon_0$&\quad\quad$8$\\
      \hline
    \hline
\end{tabular}\\
\end{table}
\end{center}

\section{Analytical solutions under optical excitation}

In this section, we present the analytical solutions of Eqs.~(\ref{SP}) and~(\ref{SA}) within the response theory, which can be applied to the case during the pump pulse. Assuming the external pump field ${\bf E}(t)={\bf E}_0\cos(\Omega{t})$ with a weak strength and neglecting all damping terms, Eq.~(\ref{SP}) can be approximated as
\begin{equation}
m_p\partial^2_t\delta{\bf P}=-\alpha_e\delta{\bf P}-3b{P^2_0}\delta{\bf P}+{\bf E}_0\cos(\Omega{t}),    
\end{equation}
which leads to the solution:
\begin{equation}
\delta{\bf P}=\frac{{\bf E}_0\cos(\Omega{t})}{\alpha_e+3bP_0^2-m_p\Omega^2}.
\end{equation}
Substituting this solution to Eq.~(\ref{SA}), one has
\begin{equation}
\partial_{t}\delta\alpha(t)=-\eta\Big[\frac{({\bf P}_0\cdot{\bf E}_0)\cos(\Omega{t})}{\alpha_e+3bP_0^2-m_p\Omega^2}+\frac{E^2_0\cos(2\Omega{t})+E_0^2}{4(\alpha_e+3bP_0^2-m_p\Omega^2)^2}\Big],  
\end{equation}
leading to the solution:
\begin{equation}\label{SSS}
\delta\alpha(t)=-\eta\Big[\frac{({\bf P}_0\cdot{\bf E}_0)\sin(\Omega{t})/\Omega}{\alpha_e+3bP_0^2-m_p\Omega^2}+\frac{E^2_0\sin(2\Omega{t})/(2\Omega)}{4(\alpha_e+3bP_0^2-m_p\Omega^2)^2}+\frac{E_0^2t}{4(\alpha_e+3bP_0^2-m_p\Omega^2)^2}\Big].  
\end{equation}
For the soft-mode dynamics in the paraelectric phase, at equilibrium $P_0=0$, and hence, the first term in Eq.~(\ref{SSS}) vanishes, with only the second-order response to the pump field remaining, i.e., the hot-phonon effect (non-oscillatory component in third term) and the soft-mode frequency doubling (the second term) phenomenon. As for the polar-mode dynamics in the ferroelectric phase, because of the equilibrium $P_{0}\ne0$, the linear response to the pump field (first term) dominates as a consequence of the breaking of the global lattice inversion symmetry by the intrinsically existing ferroelectric order, and hence, the field-induced SHG response during the pump pulse follows the pump-pulse waveform. The hot-phonon effect and soft-mode frequency doubling phenomenon that manifest in second-order-excitation regime are therefore masked. 

\begin{figure}[htb]
  {\includegraphics[width=12cm]{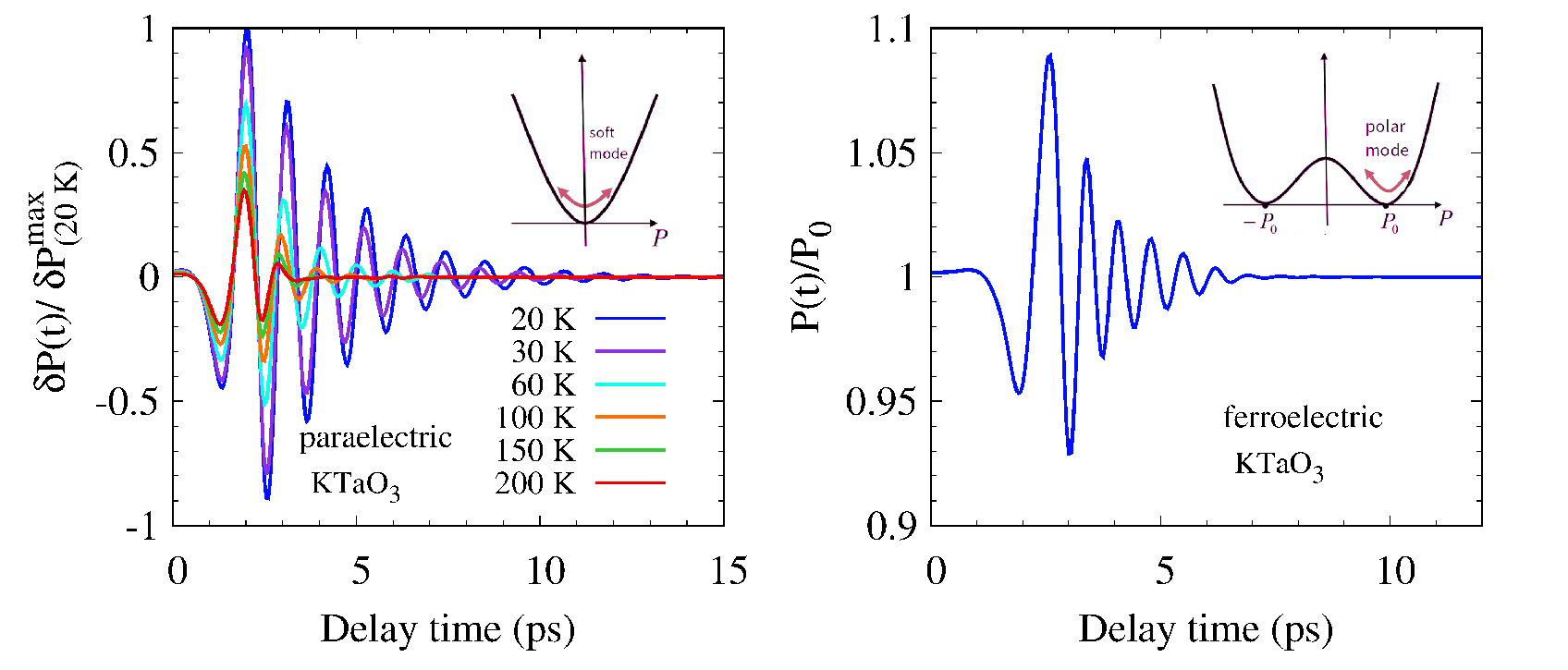}}
  \caption{{{Numerical simulation of the time-resolved polarization in paraelectric (left) and ferroelectric (right) KTaO$_3$, corresponding to the simulation presented in the main text. The insets illustrate the corresponding amplitude-mode oscillations near the energy minima in each state. }} }    
\label{Sfigyc4}
\end{figure}

\section{Dynamics of the polarization}

{{In this section, we present the corresponding numerical results of the polarization dynamics ${\bf P}(t)$, which are plotted in Fig.~\ref{Sfigyc4}. It is important to clarify that the the dynamics of $\delta\alpha(t)$ in the main text reflects the  collective (many-body) behavior of the full soft-phonon excitations, not merely the single-excitation $q \sim 0$ soft mode. The $\alpha(T)$ in Eq.~(3) of the main text is a macroscopic quantity, incorporating contributions from the entire soft-phonon branch. Due to this, the temperature dependence of the equilibrium harmonic coefficient $\alpha_e(T)$ can be described in terms of the thermal population of soft phonons, and the dynamics of $\delta\alpha(t)$ reflect the nonequilibrium hot-phonon effects.}} 

{{The $q \sim 0$ soft-mode dynamics can be directly accessed through the polarization ${\bf P}(t)$, which is given by
${\bf P} = u_{\rm sp} \left(\sum_i Q_i {\bf e}_i\right) / \Omega_{\rm cell}$,
where $u_{\rm sp}$ is the soft-mode displacement, $Q_i$ and ${\bf e}_i$ are the effective charge and eigenvector of the $i$-th ion in the unit cell (volume $\Omega_{\rm cell}$), respectively. As shown in Fig.~\ref{Sfigyc4}, in the paraelectric phase, the polarization  ${\bf P}=\delta{\bf P}$  oscillates around zero (coherently at low temperatures and incoherently at high tempratures), consistent with an uncondensed soft mode. In contrast, in the ferroelectric phase, the polarization oscillates around a nonzero value ${\bf P}_0$, i.e., ${\bf P}={\bf P}_0+\delta{\bf P}$, 
indicating a condensed soft mode, i.e., formation of ferroelectric order ${\bf P}_0={\langle}u_{\rm sp}\rangle _{\rm condensed}\times\left(\sum_i Q_i {\bf e}_i\right) / \Omega_{\rm cell}$.}} 

{{Notably, we emphasize that while the soft phonons with imaginary frequency are condensed in the ferroelectric phase, the  dynamical response of the ferroelectric polarization still corresponds to a collective amplitude excitation, polar mode, analogous to a Higgs-like mode in systems with spontaneous symmetry breaking, as illustrated in the inset of Fig.~\ref{Sfigyc4}. As a consequence, the results of polarization dynamics under resonant excitation by THz field are, in fact, naturally expected: in the ferroelectric phase, the soft mode is condensed, resulting in a spontaneous polarization corresponding to a shifted equilibrium position around which coherent oscillations occur (amplitude mode oscillations near the energy minimum). In contrast, in the paraelectric phase, the soft mode remains uncondensed, and the polarization coherently oscillates around zero displacement, reflecting the symmetric, high-symmetry phase without spontaneous polarization.}}

\section{Terahertz-pump SHG-probe measurement}

\begin{figure}[htb]
  {\includegraphics[width=12cm]{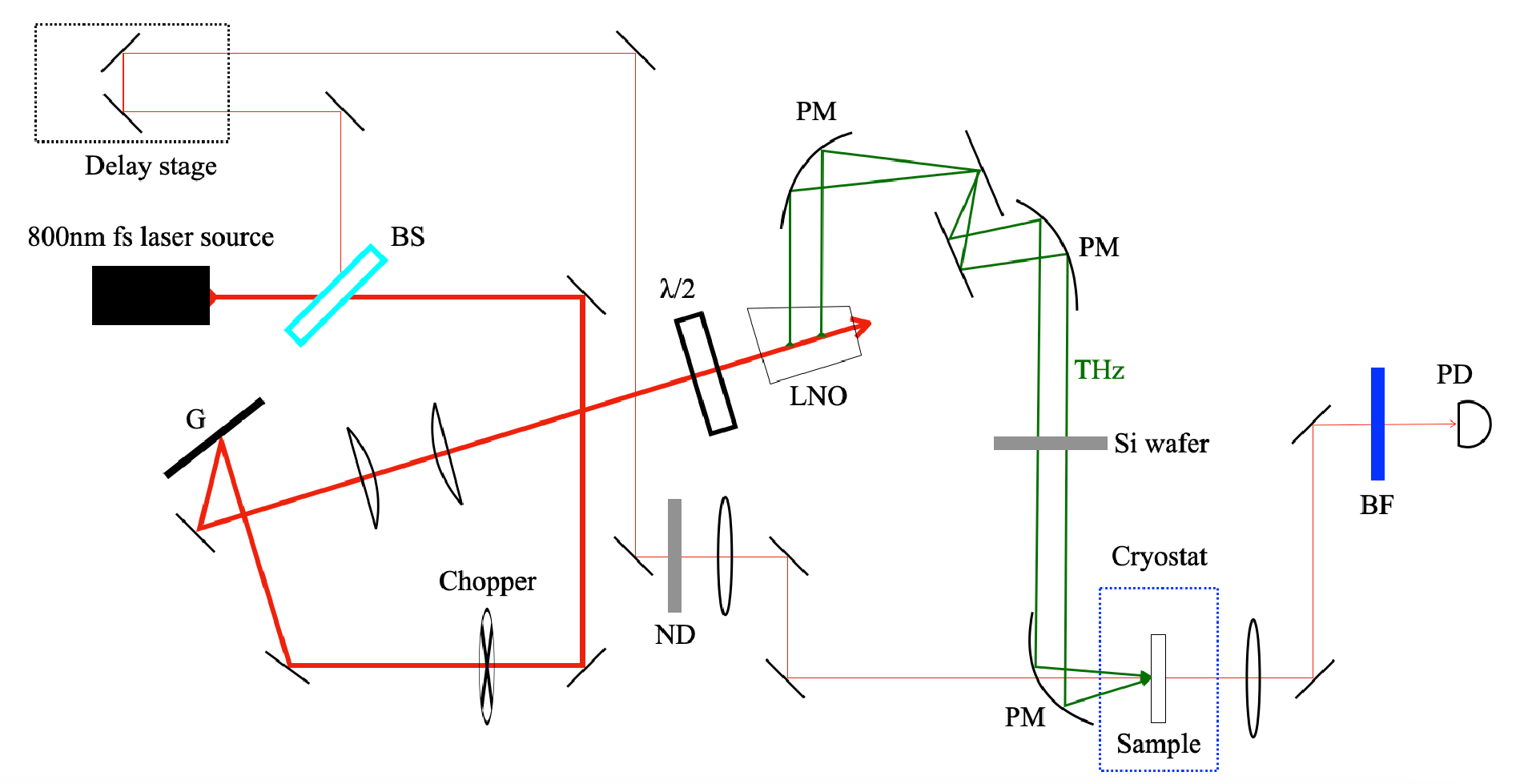}}
  \caption{{{The schematic of our experimental THz-pump SHG-probe setup. PM: parabolic mirror, BS: beam splitter, G: grating, PD: photo-diode, ND: natural density filter, BF: bandpass filter, LNO: LiNbO$_3$ crystal.}} }    
\label{THz}
\end{figure}

Figure~\ref{THz} shows the schematic of our THz-pump SHG-probe setup for the experiment. Before the THz generation, a small portion ($ < 10\%$) of the femtosecond laser is split off to serve as the probe. The intense single-cycle THz pulses are generated by optical rectification of the pulsed 800-nm fundamental beam in a LiNbO$_3$ prism with tilted-wavefront method. After the prism, the fundamental beam is screened out by a high-resistivity silicon wafer with 1~mm thickness. The pump and probe are collinear and spatially pre-overlapped at the sample position by a knife-edge method. The electric fields of the pump and probe are both aligned along the [100] direction of the sample. After the sample, the fundamental beam (800~nm) is screen out by a 400~nm bandpass filter and the SHG signal (400~nm) is detected by a GaP photodiode. 
For the temperature variation, the sample was cooled in a Janis cryostat equipped with quartz windows. The quantum paraelectric KTaO$_3$ sample used in the present study is a commercial product from MSE Supplies cut in the [100] direction. The ferroelectric KTaO$_3$ sample used in this study was prepared through an annealing process in vacuum, $10^{-5}$ Torr, with possibly tittle Sn or Fe dopants from our experimental environment, which may induce ferroelectricity similar to Nb-doped KTaO$_3$~\cite{hochli1979quantum,hochli1977quantum}.
The exact mechanism driving the transition here  is still under investigation, but it does not affect the conclusions of the present non-equilibrium study on ultrafast responses of a ferroelectric KTaO$_3$, as similar phenomena have been widely observed in THz-pump SHG-probe measurements of other ferroelectric materials~\cite{von2018probing,grishunin2017thz,miyamoto2018ultrafast}.

To extract the soft-mode frequencies at different temperatures from experimental data, as performed in the previous work~\cite{cheng2023terahertz},  we first used a single exponential relaxation function
convolving with a step function to fit the non-oscillatory component of the measured time-resolved signal. Then, we can obtain the oscillatory component by subtracting this non-oscillatory component from the original signal, and hence, the Fourier transform of the SHG oscillations [Fig.~2(a) in the main text].  The soft-mode frequencies at different temperatures are determined from the resonance peaks appearing in the range of [1.5~THz, 2.8~THz]  using a double-peak Lorentzian fitting by taking into account the presence of a primary THz-field spectrum peak around 1.8~THz.
\begin{equation}
\delta{\rm I_{SHG}^{oscillations}}(\Omega)=\frac{C_{\rm THz}}{(\Omega-\omega_{\rm THz})^2+\gamma_{\rm THz}^2}+\frac{C_{\rm sp}}{(\Omega-\omega_{\rm sp})^2+\gamma_{\rm sp}^2}.
\end{equation}
Here, $C_{\rm THz}$ and $C_{\rm sp}$ are the amplitudes for the THz-field spectrum peak and soft-mode resonance peak, respectively; $\gamma_{\rm THz}$ and $\gamma_{\rm sp}$ denote the corresponding broadening, and  $\omega_{\rm THz}$ and $\omega_{\rm sp}$ denote the corresponding center frequencies.   

\begin{figure}[htb]
  {\includegraphics[width=18.5cm]{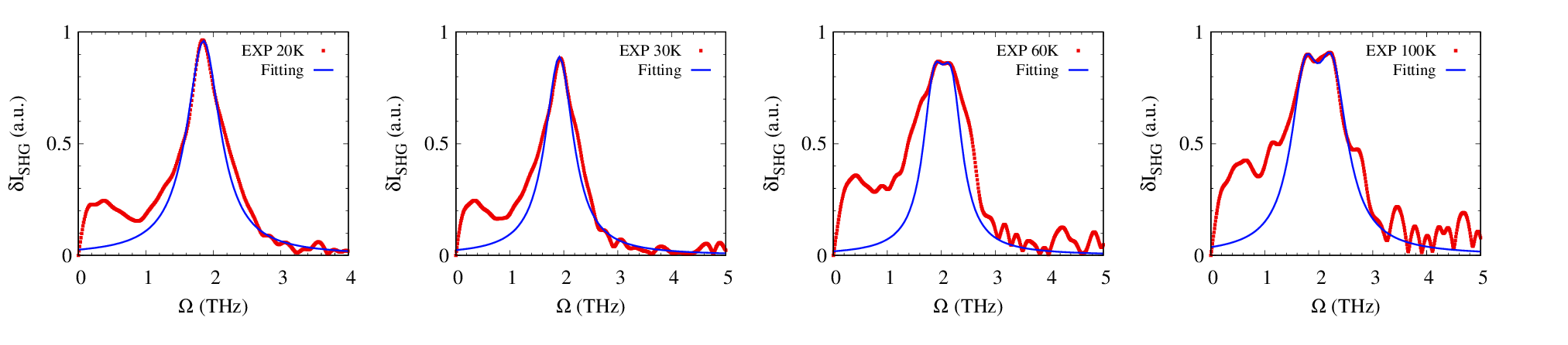}}
  \caption{{{Fit to the Fourier transform of SHG oscillations by using the double-peak Lorentzian fitting method. We find the center frequency of the THz-field spectrum peak $\omega_{\rm THz}$ is always  around $\sim$1.8~THz at different temperatures, i.e., nearly temperature-independent as it should be. }} }    
\end{figure}

{{The static SHG signal of probe field (before THz-pump excitation) is very weak, and strongly depends on the quality of the sample. For our experimental conditions (for example, at mediate 80~K, and a laser spot size around 50~$\mu$m focused by a 200 mm lens), the SHG signal read on a lock-in amplifier detected by a GaP photodiode detector is approximately 2.9~$\mu$V with 220~$\mu$W probe intensity (the noise level in our experiment is around 1~$\mu$V), confirming the negligible static contribution. Under a similar experimental condition but 500~$\mu$W probe intensity,   after the resonant excitation by a 210~kV/cm THz pump field, the reading of the SHG signal reaches up to approximately 120~$\mu$V at the peak position in the positive time delay, reflecting a significant modulation beyond the static defect-related background.  Based on these measurements, we estimate that the modification in $1/\varepsilon(t,\omega_{\rm probe})$ is enhanced by up to 8 times after THz pumping. This enhancement for high-frequency inverse dielectric constant should be scientifically reasonable for a coherent excitation by the intense THz field. However,  we emphasize that the absolute magnitude of the SHG signal before and after THz pumping can vary significantly from sample to sample and across the experimental environment conditions. This in fact complicates any attempt to draw general conclusions from the absolute intensity values alone. Therefore, we chose  normalized SHG intensity changes to perform analysis in the main text,  i.e., ${\delta}I_{\rm SHG}(t)/{\delta}I_{\rm SHG (20~K)}^{\rm max}$, where the maximum THz-induced SHG intensity change  ${\delta}I_{\rm SHG (20~K)}^{\rm max}$ is approximately 316~$\mu$V. From a theoretical standpoint, normalized responses are less sensitive to extrinsic experimental factors (e.g., sample thickness, surface roughness, laser spot alignment, or power fluctuations and in particular, the density of oxygen-vacancy defects and overall crystal quality), 
 and thus better reflects the THz-triggered intrinsic temporal dynamics of the materials, which is the key physics that we aim to understand. Therefore, presenting the normalized SHG dynamics provides a more robust and reproducible way to extract and compare the underlying physics, and  allows one to focus on the key physical mechanisms, without the need to model experimental uncertainties that are often difficult to quantify theoretically.  }}

\section{Additional Discussion}

{{Our model focuses on the dynamic variation of the dielectric environment due to the lattice background, which reflects the coherent response of the soft mode under THz pumping. As such, it is specifically designed to treat normalized THz-induced relative changes in SHG intensity, and thus at this stage, does not provide absolute SHG intensity values. To clarify, the relative change in SHG intensity is given by:
            \begin{equation}
    \frac{{\delta}I_{\rm SHG}(t)}{{\delta}I_{\rm SHG (20K)}^{\rm max}}= \frac{{\delta}(1/\varepsilon(t))}{{\delta}(1/\varepsilon)_{\rm (20K)}^{\rm max}}=\frac{\delta\alpha(t)}{\delta\alpha^{\rm max}_{\rm (20K)}}.
\end{equation}    
One of the key features of the current theoretical framework is its relative simplicity and yet is able to quantify the normalized THz-induced SHG changes. A rigorous formulation of absolute SHG intensities  ${\rm I}_{\rm SHG}=\frac{1}{2\varepsilon}{p^{(2\omega)}_{\rm probe}p^{(2\omega)}_{\rm probe}}$ requires a detailed understanding of the SHG origin before THz pumping, specifically the calculation of the probe-field SHG polarization  $p^{(2\omega)}_{\rm probe}$. As mentioned above, the origin of this finite polarization in quantum paraelectrics is most likely linked to inevitable oxygen vacancies. More microscopically, it appears to arise from two-photon inter-band transitions of electrons, mediated by electronic defect states associated with these oxygen vacancies.  To calculate this transition requires density functional theory (DFT) calculations to evaluate oxygen-mediated two-photon inter-band transition matrix elements and the development of a non-equilibrium electronic transition model (e.g., optical Bloch equations in semiconductors).  This involves much more complex contributions beyond the scope of our current framework.  
}}

{{Moreover, the THz-induced SHG intensity modulation in our model is governed by the change in the high-frequency inverse dielectric function.  
 This physical phenomenon, where low-frequency structural modulation impacts high-frequency optical response, is widely reported in nonlinear phononics and light–matter coupling in polar dielectrics. Specifically, in most of pump-probe measurements,  
 although the SHG probe operates at optical frequencies, the low-frequency THz pump can still significantly influence the optical-frequency  dielectric function. Physically, this is because that relative to the high optical frequencies of the probe field, the THz pump can be regarded as quasi-static during the SHG process. This quasi-static lattice deformation modifies the local crystal field environment and electronic polarizability, which in turn affects $\varepsilon(\omega_\mathrm{probe})$ even though the pump and probe operate at vastly different frequencies.  In simpler terms,  `the low-frequency' THz pump essentially sets the stage for how the electrons in the system respond to the high-frequency probe field.}} 

 {{Within our framework, we treat this effect via the soft-phonon dynamics $\alpha(t)$ (driven by the THz field) and assumption of $1/\varepsilon(t,\omega_{\rm probe}) \propto \alpha(t)$, which encodes the THz-driven coherent phonon dynamics that modulate the effective high-frequency response. While this captures the essential impact of THz-driven lattice dynamics on the optical response, it does not explicitly resolve the full frequency dependence of the dielectric constant. A rigorous calculation of the full frequency-dependent dielectric response under nonequilibrium conditions (i.e., nonlinear optics) would require advanced first-principles methods (such as time-dependent density functional theory or many-body perturbation theory) that incorporate both pump-induced nonequilibrium lattice and electronic contributions and in particular their coupling ({\sl electron-phonon interaction}) to the dielectric environment across a broad frequency range of probe field.  Such calculations are beyond the scope of the present work. }}
